\documentstyle[12pt]{article}

\def\gsim{\raisebox{-4pt}{$\,\stackrel{\textstyle{>}}{\sim}\,$}}
\begin{document}
\begin{flushright}
\baselineskip=12pt
CERN--TH/96--306\\
DOE/ER/40717--35\\
CTP-TAMU-53/96\\
ACT-16/96\\
\tt hep-ph/9610470
\end{flushright}

\begin{center}
\vglue 1.5cm
{\Large\bf Analysis of LEP Constraints on Supersymmetric Models with a
Light Gravitino}
\vglue 1.5cm
{\Large John Ellis $^1$, Jorge L. Lopez$^2$, and D.V. Nanopoulos$^{3,4}$}
\vglue 1cm
\begin{flushleft}
$^1$Theory Division, CERN, 1211 Geneva 23, Switzerland\\
$^2$Bonner Nuclear Lab, Department of Physics, Rice University\\ 6100 Main
Street, Houston, TX 77005, USA\\
$^3$Center for Theoretical Physics, Department of Physics, Texas A\&M
University\\ College Station, TX 77843--4242, USA\\
$^4$Astroparticle Physics Group, Houston Advanced Research Center (HARC)\\
The Mitchell Campus, The Woodlands, TX 77381, USA\\
\end{flushleft}
\end{center}

\vglue 1cm
\begin{abstract}
We propose an analysis of LEP constraints on radiative neutralino
decays into a light gravitino, based on the plane of the Higgs mixing
parameter $\mu$ and the $SU(2)$ gaugino mass $M_2$. The preliminary LEP 2W
constraints in the $(\mu, M_2)$ plane are considerably stronger than for
supersymmetric models in which the lightest neutralino is stable. A significant
portion of the parameter space in which chargino or selectron decay into a
final state containing a light gravitino could provide an interpretation of the CDF $e^+e^-\gamma\gamma +{\rm E_{T,miss}}$ event can now excluded by the
preliminary LEP 2W data.
\end{abstract}

\vspace{1cm}
\begin{flushleft}
\baselineskip=12pt
October 1996\\
\end{flushleft}
\newpage
\setcounter{page}{1}
\pagestyle{plain}
\baselineskip=14pt

There are three generic phenomenological scenarios for the
lightest supersymmetric particle (LSP). Either (i) $R$ parity is violated
and the LSP is unstable, or $R$ parity is an exact symmetry and the LSP is
stable, in which case it is presumably neutral and at most only
weakly interacting~\cite{stable}, and may be either (ii) the lightest
neutralino
$\chi$, or (iii) some still lighter sparticle such as the gravitino
$\widetilde G$. Most phenomenological studies have been in the context of
the second scenario, in which the $\chi$ is the stable LSP~\cite{stable},
though $R$-violating models (i) have also been studied~\cite{RV}.
$R$-conserving models (iii), in which the LSP is not the $\chi$, have been
around for some time~\cite{Fayet,EEN}, but have only recently attracted
considerable attention \cite{KaneDine,Gravitino}. This has been revived by the
CDF report of a single $e^+ e^- \gamma\gamma+{\rm E_{T,miss}}$ event
\cite{Park}, but is an interesting generic possibility in its own right,
independent of the CDF report. In our view, this class of models (iii) should
be studied in greater depth: in particular, the LEP constraints on this
scenario should be explored just as thoroughly as for  $R$-violating models
(i), and models (ii) in which the $\chi$ is the stable LSP.

Several discussions of the implications of class (iii) models for
LEP phenomenology have already appeared \cite{KaneDine,Gravitino}. Our purpose
here is to propose a general analysis strategy for these models which is
adapted from those already used for the other classes of models, and might be
suitable for adoption in experimental analyses. Discussions of charginos and
neutralinos in models of classes (i)~\cite{RVALEPH} and (ii) usually start
from an investigation of the $(\mu, M_2)$ plane~\cite{ALEPH,EFOS}, where $\mu$
is the familiar Higgs superpotential mixing parameter, and $M_2$ is
the $SU(2)$ gaugino mass. As in most analyses of class (ii) models,
we assume $SU(2):U(1)$ gaugino universality at the supersymmetric
GUT scale, so that $M_1=(\alpha_1/\alpha_2)M_2$. It is known~\cite{JN} that
this does not affect greatly the $(\mu, M_2)$ analysis in class (ii)
models, and we do not expect it to be a sensitive assumption here,
either. Auxiliary parameters in a $(\mu, M_2)$ analysis are the ratio
tan$\beta$ of Higgs v.e.v.'s, and the masses of the sleptons $\tilde \ell$
and $\tilde \nu$, which affect the cross sections at LEP for associated
production of pairs of neutralinos $\chi^0_i \chi^0_j$ and pair production of the lighter chargino $\chi^+ \chi^-$, respectively. The $(\mu, M_2)$ plane
is well suited for exposing the domain of parameter space in which the CDF
event~\cite{Park} is interpreted as chargino pair production, followed by
$\chi^{\pm}$ decay into a pair of $e + \nu + \gamma + {\widetilde G}$ final
states.

It is also usual to analyze LEP constraints on slepton production in class (ii)
models using the plane of the parameters $(m_{\tilde \ell}, m_{\chi})$,
which are related simply to underlying supergravity parameters
$(m_0, m_{1/2})$~\cite{EFOS}. This plane can also usefully be analyzed in
class (iii) models, taking into account the $(\mu, M_2)$ analysis proposed
above~\footnote{We note that it is not possible to carry directly over to
class (iii) models the results of the slepton searches in class (ii) models, since most of the latter veto events containing photons.}.
This type of analysis is useful for comparison with the selectron-pair
production interpretation of the CDF event~\cite{Park}.

In this paper, we apply these analysis steps to the preliminary data
recently announced by the four LEP collaborations \cite{Oct8}, obtained during
the recent run of LEP just above the $W^+ W^-$ threshold at $161$ GeV,
which we term LEP 2W. After applying mild experimental cuts~\cite{Dann},
no acoplanar $\gamma\gamma+{\rm E_{miss}}$ events were found by the
DELPHI, ALEPH and OPAL collaborations, whereas L3 have reported 2 events.
On the basis of their absence of events, the DELPHI, ALEPH and OPAL
collaborations have quoted preliminary upper limits on the cross section
$\sigma_{\gamma\gamma}$ for such events of $\sim 0.5, 0.4$ and $0.4$
pb~\cite{Oct8}, respectively. As a basis for our discussion, we interpret
these as a combined LEP 2W upper limit $\sigma_{\gamma\gamma} < 0.2$ pb. Our
qualitative conclusions will be insensitive to the precise numerical
value of this upper limit~\footnote{Although we have in mind a no-scale
supergravity model with a light gravitino~\cite{EEN,Gravitino}, the
essential features of our approach are also applicable to gauge-mediated models~\cite{KaneDine}, and to models in which the lightest supersymmetric particle is an axino \cite{Hisano}, as long as the LSP is very much lighter than the $\chi$.}.

\begin{figure}[p]
\vspace{6in}
\includegraphics{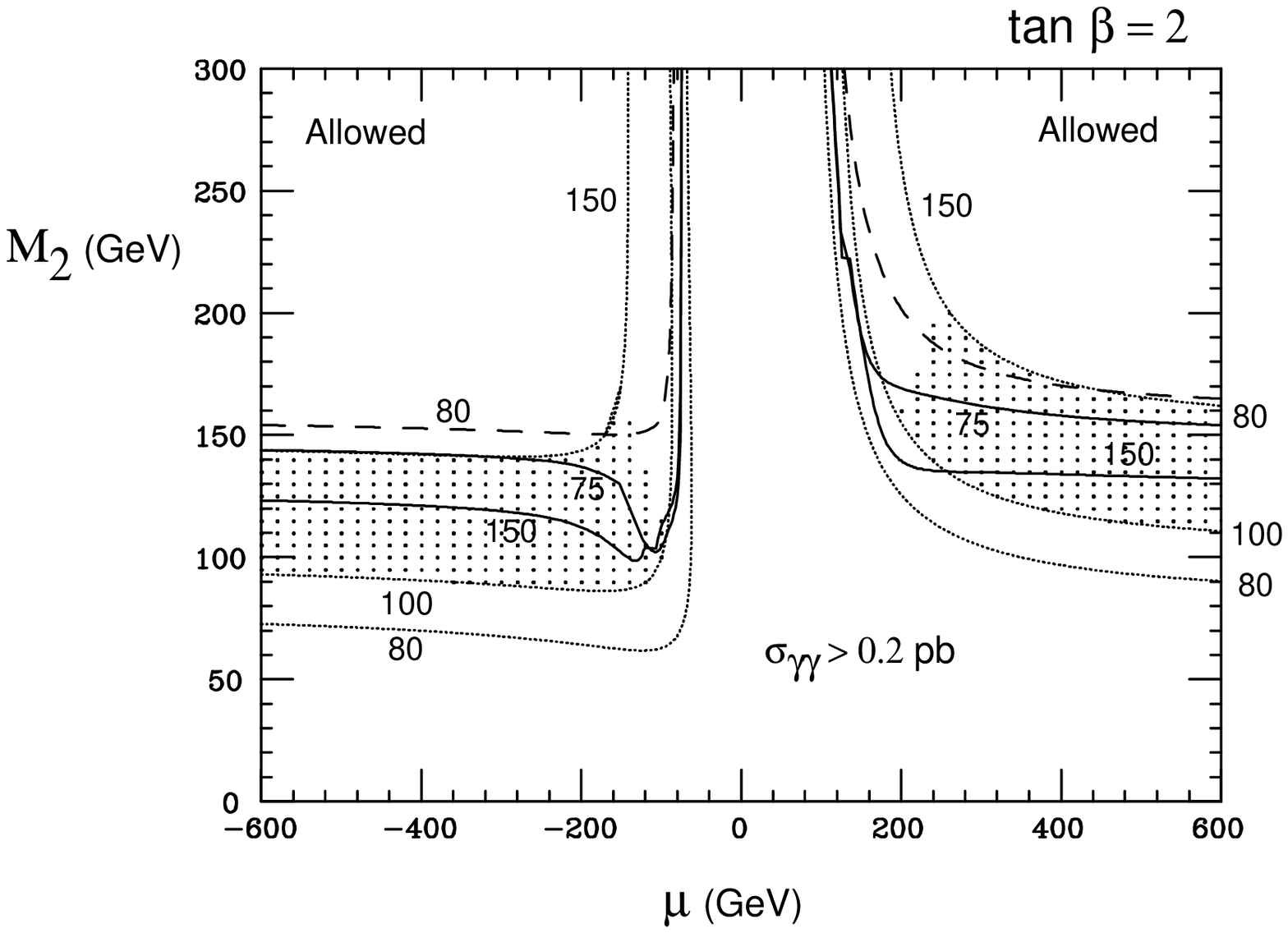}
\caption{The $(\mu, M_2)$ plane for $\tan\beta=2$, indicating the
$\sigma_{\gamma\gamma} = 0.2$ pb contour for $m_{\tilde e}=75,150$ GeV (solid
lines). Domains below and between the two arms of the solid lines are excluded
by our interpretation of 
the LEP 2W data~[12]. Also indicated are the contours of
$m_{\chi} = 80$ GeV (dashed lines), and $m_{\chi^\pm}= 80, 100, 150$ GeV
(dotted lines). The chargino interpretation of the  CDF event~[7] requires
$m_{\chi^\pm}\approx(100-150)$ GeV and  $m_{\chi} < 0.6 m_{\chi^\pm}$
(dotted regions).}
\label{fig:M2mu2}
\end{figure}

\begin{figure}[p]
\vspace{6in}
\includegraphics{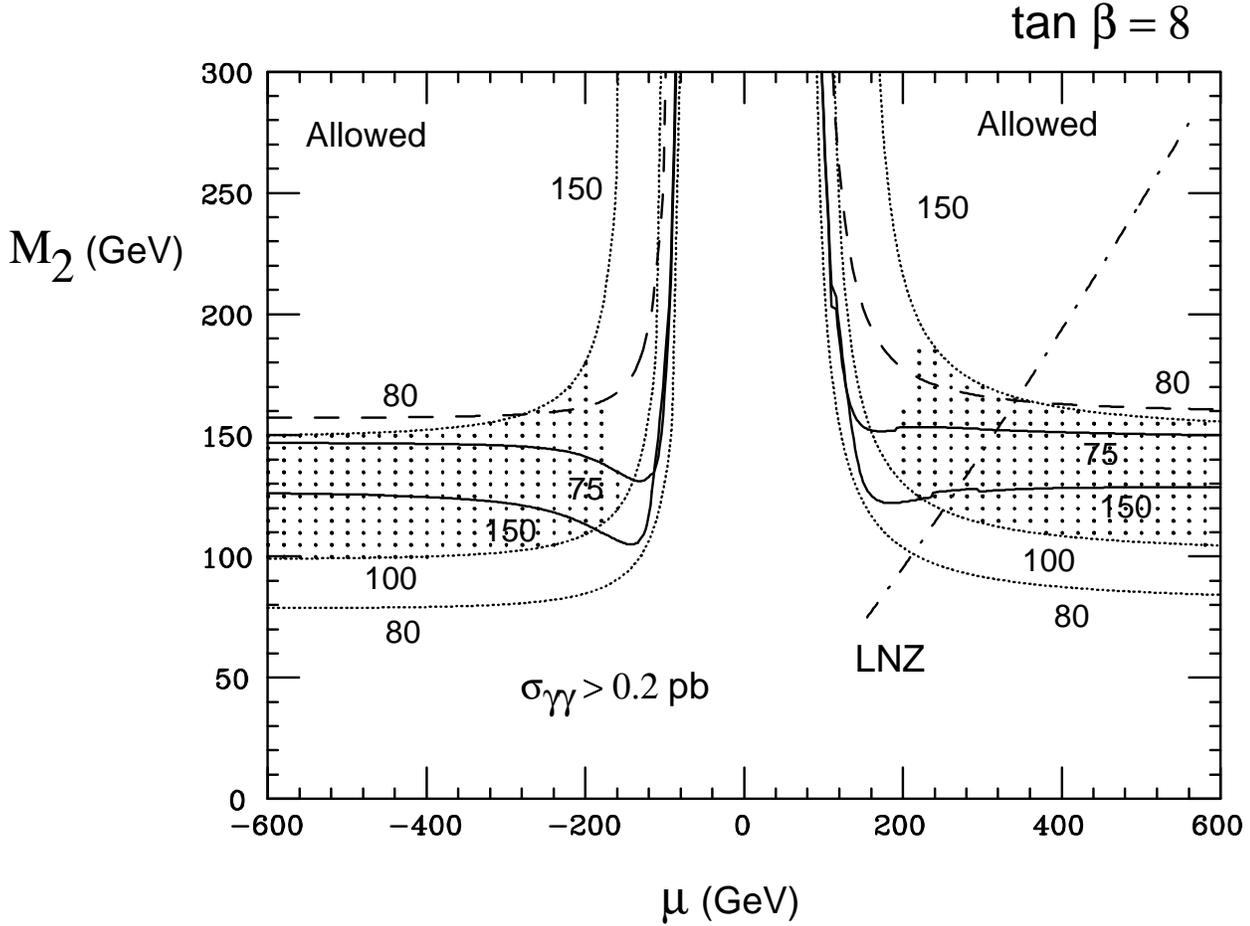}
\caption{The $(\mu, M_2)$ plane for $\tan\beta=8$, indicating the
$\sigma_{\gamma\gamma} = 0.2$ pb contour for $m_{\tilde e}=75,150$ GeV (solid
lines). Domains below and between the two arms of the solid lines are excluded
by our interpretation of
the LEP 2W data~[12]. Also indicated are the contours of
$m_{\chi} = 80$ GeV (dashed lines), and $m_{\chi^\pm}= 80, 100, 150$ GeV
(dotted lines). The chargino interpretation of the  CDF event~[7] requires
$m_{\chi^\pm}\approx(100-150)$ GeV and  $m_{\chi} < 0.6 m_{\chi^\pm}$
(dotted region). We also indicate the ray singled out in the model of Ref.~[6]
(dot-dashed line).}
\label{fig:M2mu8}
\end{figure}

We start by exploring the $(\mu,M_2)$ plane for the representative choices
tan$\beta = 2, 8$ shown in Figs.~\ref{fig:M2mu2},~\ref{fig:M2mu8}. The
process which gives the most stringent constraints in this plane is $e^+ e^-
\rightarrow \chi \chi$ (see also~\cite{RVALEPH}), which depends on the
selectron mass $m_{\tilde e}$ (assumed here to be degenerate: $m_{\tilde e_L}=m_{\tilde e_R}$), followed by $\chi\to\gamma + {\widetilde G}$ decay. The associated production of $\chi$ and $\chi_2$, followed by $\chi_2 \rightarrow \chi + \nu + {\bar\nu}$ and $\chi \rightarrow \gamma + {\tilde G}$ decays, may also contribute to $\sigma_{\gamma\gamma}$ \cite{Gravitino}, so it is conservative to retain just the $\chi \chi$ production process. The solid lines correspond to $\sigma_{\gamma\gamma} = 0.2$ pb for the two limiting values
$m_{\tilde e} = 75, 150$ GeV, the lower value being close to the
LEP 2W kinematic limit, and the higher value corresponding to the highest
selectron mass consistent with the selectron pair-production interpretation
of the CDF event~\footnote{Note, however, that these choices are not
conservative, in the sense that the $\chi \chi$ cross section would be
smaller for $m_{\tilde e}$ beyond this CDF-motivated range.}. As
an example of the case of non-degenerate masses, for $m_{\tilde e_R}=75\,{\rm GeV}$ and $m_{\tilde e_L}=150\,{\rm GeV}$ we obtain a line between the two solid lines in Figs.~\ref{fig:M2mu2},~\ref{fig:M2mu8}. We recall that left- and
right-handed selectrons couple differently to neutralinos, depending on the
neutralino composition. In the limit $|\mu| \gg M_2$, where the lightest
neutralino is asymptotically a pure $\tilde B$, its coupling to the
${\tilde e}_R$ is larger than that to the ${\tilde e}_L$, so the common
mass we use here would be closer to $m_{{\tilde e}_R}$ in a model with
non-degenerate masses.

The domains of parameter space below and between the two arms
of the solid lines are excluded by our interpretation of the preliminary
LEP 2W data. The dashed lines are the contours where $m_{\chi} = 80$
GeV, which was the kinematic limit for LEP 2W, which is approached quite
closely if $m_{\tilde e} = 75$ GeV. The dotted lines are contours of
the chargino mass $m_{\chi^\pm} = 80, 100, 150$ GeV. They represent,
respectively, the kinematic limit of LEP 2W~\footnote{We note in passing that
LEP 2W searches are probably sensitive to the $e^+e^- \to \chi^+ \chi^-$
process for most values of $m_{\chi^{\pm}}$ up to this limit. In contrast to
the conventional stable-neutralino scenario, where experimental sensitivity is
lost when $m_{\chi^{\pm}} > m_{\tilde \nu} > m_{\chi^{\pm}} - 10$ GeV,
here there will always be a signature of two energetic photons.}, the
lower limit on $m_{\chi^\pm}$ in the chargino interpretation of the CDF
event, and an estimate of the upper limit on $m_{\chi^{\pm}}$ in this
interpretation~\footnote{Note, however, that the possibility of models
with $m_{\chi^{\pm}}$ up to 200 GeV has also been considered~\cite{KaneDine}, though the resulting production cross section at the Tevatron may then become
rather small.}. The dotted region is that in which the chargino interpretation
may be valid, with the constraint $m_{\chi} < 0.6 m_{\chi^\pm}$ also applied
\cite{Gravitino}. Most models capable of fitting the CDF event in fact
have $m_{\chi} < 0.5 m_{\chi^\pm}$: applying this constraint would bound
the dotted region further away from the $\mu = 0$ line, reducing the scope for a model to lie above the $m_{\tilde e} = 75$ GeV solid line.

It is immediately apparent from Figs.~\ref{fig:M2mu2},~\ref{fig:M2mu8}
that the LEP 2W bounds in the class (iii) radiative decay framework discussed
here are much stronger than those in the conventional stable-neutralino
scenario of class (ii), at least in the region of the $(\mu, M_2)$ plane where
the lightest neutralino has a predominant gaugino component. In the limit
$|\mu| \gg M_2$, where the lightest neutralino is almost a pure $U(1)$
gaugino $\widetilde B$, the LEP 2W lower limit on $M_2$ may be almost a
factor two higher in class (iii) models than in class (ii) models. In
particular, the direct lower limit $m_{\chi^\pm} > 80$ GeV may be
improved to $m_{\chi^\pm} > 150$ GeV for small $m_{\tilde e}$.

This observation implies that a significant fraction of the range of
$m_{\chi^\pm}$ in which the chargino interpretation of the CDF event is
tenable may be excluded by the preliminary LEP 2W data, if $m_{\tilde
e}$ is not very large. Looking in more detail at
Figs.~\ref{fig:M2mu2},~\ref{fig:M2mu8}, we see that the LEP 2W bounds may
be least restrictive for models in which $m_{\chi}$ is close to the upper limit
of $0.6 m_{\chi^\pm}$ or for selectron masses that are not too small. It so
happens that the specific no-scale supergravity model studied in
\cite{Gravitino} appears just in this particular region, for tan$\beta \sim 8$,
as indicated by the dot-dashed line in Fig.~\ref{fig:M2mu8}. For reference, in
this model the selectron masses when entering (leaving) the chargino region
({\em i.e.}, for $m_{\chi^\pm}=100\,(150)\,{\rm GeV}$) are $m_{\tilde
e_R}=88\,(115)\,{\rm GeV}$ and $m_{\tilde e_L}=133\,(181)\,{\rm GeV}$. Taking
these selectron mass variations into account, one may conclude that
$m_\chi>70\,{\rm GeV}$ is required, which corresponds to excluding
approximately the first half of the portion of the dot-dashed curve that
intercepts the dotted region.

\begin{figure}[p]
\vspace{6in}
\includegraphics{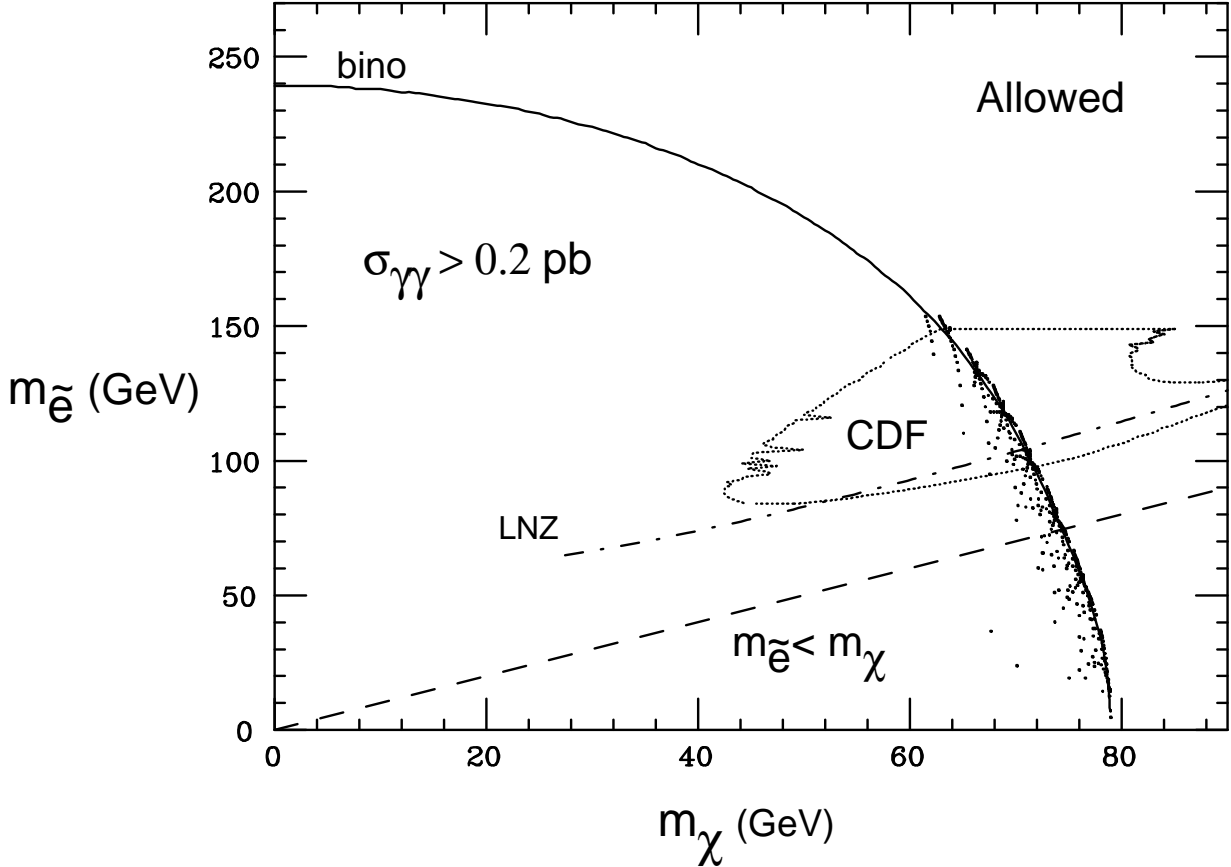}
\caption{The $(m_{\chi}, m_{\tilde e})$ plane, showing the region
where $\sigma(e^+e^- \to \chi \chi) > 0.2$ pb, so that
$\sigma_{\gamma\gamma}$ presumably exceeds the limit imposed by the preliminary
LEP 2W data~[12]. The solid line applies to the limit where $\chi$
is a pure $\widetilde B$, namely when $|\mu| \gg M_2$. The dots represent
models with $m_{\chi} < 80$ GeV and $m_{\chi^{\pm}}>125$ GeV, for which
selectron production is likely to be more significant for searches at the
Tevatron. The region where the kinematics of the CDF event~[7] are
compatible with this selectron interpretation is delineated by the dotted
lines. We also indicate the line singled out in the model of Ref.~[6]
(dot-dashed line) and the region where $m_{\tilde e} <
m_{\chi}$ (dashed line).}
\label{fig:selchi}
\end{figure}

We now turn to the analysis of the $(m_{\chi}, m_{\tilde e})$ plane. To
simplify this analysis initially, we consider the limit of large $|\mu|$,
where $\chi$ is asymptotically a pure $\widetilde B$ state, and $\tan\beta$
becomes an irrelevant parameter. In this limiting case, we find the
contour shown as a solid line in Fig.~\ref{fig:selchi}, where
$\sigma(e^+ e^- \to \chi \chi) = 0.2$ pb, the upper limit on
$\sigma_{\gamma\gamma}$ that we infer from the preliminary LEP 2W data. 
The large-$|\mu|$ asymptotic limits of the solid lines for the 
values $m_{\tilde e} = 75, 150$ GeV in
Figs.~\ref{fig:M2mu2},~\ref{fig:M2mu8} may be read from this plot.

The region of the $(m_{\chi},m_{\tilde e})$ plane that is consistent with the
kinematics of the selectron interpretation of the CDF event \cite{Gravitino} is
delineated by the dotted lines in Fig.~\ref{fig:selchi}. We see that a
significant fraction of this region is excluded by our interpretation of the
preliminary LEP 2W data, if one is in the $\widetilde B$ limit: $|\mu| \gg
M_2$. To assess the significance of the inferred LEP 2W limit for the selectron
interpretation away from this limit, we have generated a set of 
$\tan\beta=2,8$ models with $m_{\chi} < 80$ GeV (so as to be accessible at LEP 2W) and $m_{\chi^{\pm}} >125$ GeV (so that the dominant source of events for CDF is selectron-pair production), but no other {\em a priori} selection of $\mu$ or $M_2$. For each of these models, we have then found the value
of $m_{\tilde e}$ that yields $\sigma_{\gamma\gamma}=0.2$ pb. These models are
shown as dots in Fig.~\ref{fig:selchi}. We see that they cluster relatively
close to the $\widetilde B$ line. Thus, a non-negligible fraction of the
parameter space for the selectron interpretation of the CDF event is also
explored by LEP 2W, even away from the $\widetilde B$ limit. 

The selectron interpretation is therefore significantly constrained by the preliminary LEP 2W data, as we already showed to be the case for the chargino interpretation. Models compatible with both the preliminary LEP 2W data and the selectron interpretation of the CDF event are required to have $m_{\tilde e} > 95$ GeV, beyond the reach of future LEP 2 upgrades. In the case of the specific model in Ref.~\cite{Gravitino}, the correlation between the (right-handed) selectron and neutralino masses is indicated by the dot-dashed line, most of which lies within the region consistent with the kinematics of the CDF event in the selectron interpretation. The LEP 2W constraints in this model require $m_\chi>70\,{\rm GeV}$ and $m_{\tilde e} > 105$ GeV, bounded from below by a point which lies very close to the pure $\widetilde B$ line.

Over the next couple of years, the LEP beam energy will be increased in
steps up to about $96$ GeV. This will enable the sensitivity in $m_{\chi}$
to be extended to about $95$ GeV. As can be seen in Fig.~\ref{fig:selchi}, this
should be sufficient to explore essentially all the domain of the $(m_{\chi},
m_{\tilde e})$ region compatible with the selectron interpretation of the CDF
event, at least in the large-$|\mu|$ limit (and certainly in the model of
Ref.~\cite{Gravitino}). Turning back to Figs.~\ref{fig:M2mu2},~\ref{fig:M2mu8},
we see that this is not necessarily the case for the chargino interpretation.
If $m_{\tilde e}$ is large ({\em i.e.}, $m_{\tilde e}\gsim200\,{\rm GeV}$),
and/or if $m_{\chi} \rightarrow 0.6m_{\chi^{\pm}}$, there are regions of
parameter space that will not be accessible to LEP 2, even at its maximum
energy.

The main purpose of this paper has not been to consider the
possible implications of the present preliminary LEP 2W data,
or possible future LEP 2 data, for any specific light-gravitino
model, whether or not it is motivated by the CDF event.
Our objective has rather been to indicate how one may analyze experimental
constraints on such models, using an approach adapted from
previous analyses of models in which the lightest neutralino is stable.
As we have emphasized, the LEP constraints on the unstable-neutralino
models may be even stronger than those on stable-neutralino models,
because the presence of a pair of energetic photons provides an
additional signature that enables, in particular, the process
$e^+ e^- \to \chi \chi$ -- which has the lowest threshold
of any supersymmetric process -- to be observed. We believe that
future analyses by the LEP collaborations will enable a large
fraction of the parameter space of such models to be explored.

\section*{Acknowledgements}
J.E. thanks Michael Schmitt for useful discussions.
The work of J.L. has been supported in part by DOE grant
DE-FG05-93-ER-40717, and that of
D.V.N. has been supported in part by DOE grant DE-FG05-91-ER-40633.

\clearpage

\clearpage

\clearpage

\end{document}